\documentclass[12pt]{article}
\usepackage{amsmath,amssymb}
\usepackage{latexsym,graphicx,epsfig}
\newcommand{\be}{\begin{equation}}
\newcommand{\ee}{\end{equation}}
\newcommand{\bes}{\begin{subequations}}
\newcommand{\ees}{\end{subequations}}
\newcommand{\bea}{\begin{eqnarray}}
\newcommand{\eea}{\end{eqnarray}}
\newcommand{\bear}{\begin{equation}\begin{array}}
\newcommand{\eear}[1]{\end{array}\label{#1}\end{equation}}
\def\ba{$$\begin{array}}
\def\ea{\end{array}$$}
\def\bra{$\begin{array}}
 \def\era{\end{array}$}

\newcommand{\fr}[2]{\dfrac{{ #1}}{{ #2}}}

\newcommand{\f}{\phi}
\newcommand{\fd}{\phi^\dagger}
\newcommand{\zz}[4]{Z^{#1}{}_{#2}{}^{#3}{}_{#4}}
\newcommand{\zzz}[4]{{{{Z^{#1}}_{#2}}^{#3}}_{#4}}

\newcommand{\lr}[1]{ \langle #1 \rangle}

{\end{list}}
\newcounter{enumct}

\begin{document}
\renewcommand{\tilde}{\widetilde}%[1]

\date{}
\title{Towards thermal evolution of the most general 2HDM}
\author{I.~P.~Ivanov
\\
  {\small IFPA, Universit\'{e} de Li\`{e}ge, All\'{e}e du 6 Ao\^{u}t 17, b\^{a}timent B5a, 4000 Li\`{e}ge, Belgium}\\
  {\small and }\\
  {\small Sobolev Institute of Mathematics, Koptyug avenue 4, 630090, Novosibirsk, Russia}\\
  }
 \maketitle

%%%%%%%%%%%%%%%%%%%%%%%%%%%%%%%%%%%%%%%%%%%%%%%%%%%%%%%%%%%%%%%%%%%%%
\begin{abstract}
We consider the scalar sector of the most general renormalizable two-Higgs-doublet model at non-zero temperature.
We calculate the largest finite temperature corrections to the free-energy density
and study thermal evolution of the ground state.
Within the approximation chosen, we establish all possible sequences of
thermal phase transitions and study their relation with the symmetries of the model.
We show, in particular, that a charge-breaking or a $CP$-violating vacuum
can arise at intermediate stages of thermal evolution,
and that the first-order phase transition is associated with
a discrete symmetry of the potential, but not of the entire scalar lagrangian.
\end{abstract}


\maketitle

\section{Introduction}

\subsection{Why general 2HDM?}

The two-Higgs-doublet model (2HDM) is one of the simplest beyond-the-Standard-Model
extensions of the Higgs mechanism of the electroweak symmetry
breaking, \cite{TDLee,CPNSh}. In this model one introduces two
doublets of Higgs fields $\phi_1$ and $\phi_2$, which interact
with the matter fields and also self-interact via an appropriate
Higgs potential.

When constructing the Higgs potential of 2HDM,
one encounters many free parameters, which are not constrained by experiment.
Extensive studies conducted over past decades have shown that playing even with a small subset
of these free parameters one can get a rich spectrum of models with different
and quite remarkable phenomenological consequences (see examples in \cite{CPNSh}).

Still, there is something to worry about in this zoo of particular versions of 2HDM.
First, it has been noted that in certain cases seemingly different 2HDM models 
lead to very similar, and sometimes identical, physics.
These sporadic observations evolved later into the concept of
reparametrization-invariance (or Higgs-basis-invariance) of the model.
Understanding began to grow that it is the set of {\em basis-invariant features} of the theory,
and not the entire set of free parameters, that really shapes the phenomenology.
Papers \cite{basis-inv,nishi} were driven precisely
by the search for an efficient basis-invariant description of the general 2HDM.
However, a method how to extract these basis-invariant features from the conventional
parametrization of the potential remained elusive for a long time; search for basis-invariant conditions
for the explicit $CP$-violation conducted in \cite{basis-inv-CP} nicely illustrates this kind of difficulties.

Second, starting from the very first paper on this subject, \cite{TDLee},
there has always been a general understanding that extra symmetries of the 2HDM lagrangian
have a profound effect on the resulting phenomenology.
Indeed, majority of papers on 2HDM were centered around the concepts of $CP$-violation
in the scalar sector, or additional $Z_2$ or $U(1)$ symmetries of the potential.
It has been noted that such symmetries can be implemented in seemingly different
ways, still leading to similar physics,
see e.g. recent analysis of generalized $CP$-transformations in \cite{GCP}.
However, a complete list of possible symmetries and their
phenomenological consequences was very difficult to establish in a traditional approach
and was missing until very recently.

These issues left many researchers wondering if the most general treatment of 2HDM can be carried through at all.
If that were possible, one could generate the exhaustive list of possibilities offered by two Higgs doublets,
gain a deeper understanding of various particular versions of 2HDM
and clarify relations among them. This would also become a useful playground for model-builders,
who wanted to construct even more complicated Higgs sectors with predefined properties.

In addition, one should not forget the Minimal Supersymmetric Standard Model (MSSM), which has always
been an important motivation for the study of 2HDM.
Although the Higgs potential in MSSM is very constrained by the supersymmetry,
with several relations among its parameters,
supersymmetry-breaking loop corrections will lift them.
Thus, as stated in \cite{basis-inv}, the Higgs sector
of the (broken) MSSM can be described by an effective field theory
consisting of the most general 2HDM.

Thus, the most general 2HDM with no {\em a priori} restriction on its free parameters
should be viewed as a useful {\em tool} rather than a viable approximation to the reality,
and it is definitely worth studying in as much detail as possible.

\subsection{Problems with the general 2HDM}

The problem with the general 2HDM is that it cannot be worked out with straightforward algebra.
The obstacle arises at the very first step: when minimizing the Higgs potential,
one arrives at algebraic equations of sixth order, which cannot be solved in the general case.
In this situation, any method that would give any non-trivial insight into the most general 2HDM
is very welcome.

Recently, in \cite{Ivanov0,Ivanov1,Ivanov2} a geometric approach was developed that allows one
to analyze many features of the most general 2HDM circumventing the algebraic difficulty.
This approach led to proof of various coexistence theorems, classification of the number and
symmetry properties of the minima of the potential, and eventually
to the complete description of the phase diagram of the scalar sector of 2HDM.
All this was done without explicitly computing the exact position of the global minimum
of the potential.

Those results constituted the very first step towards the phenomenology of the most general 2HDM.
In this paper we make another step: we analyze how the ground state of the model changes
at non-zero temperature.
We calculate thermal corrections to the Higgs potential in the lowest non-trivial approximation
(first order in $\lambda_i$ and enhanced by $T^2$-factor), and study the thermal phase transitions
that take place. It is remarkable that this, too, can be done without explicitly
minimizing the free-energy density.

We stress once again that it is not our aim here to provide a phenomenologically viable
approximations of reality. What we do here is we develop
the only formalism available so far that works for the most general 2HDM,
hoping that its results will be used later
in more phenomenology-oriented studies.

Thermal field theory is a very developed branch of QFT.
Thermal phase transitions in multicomponent scalar field theories
have been, of course, studied previously, see e.g. \cite{Linde,Weinberg,Dvali95,Dvali96,Turok,2HDMthermal}.
A detailed analysis of the Higgs potential at finite temperature in 2HDM was conducted
for example in \cite{Turok,2HDMthermal}.
We also draw reader's attention to the paper \cite{Gufan} where an attempt was made to describe
the phase diagram of the 2HDM and study the phase transitions possible,
which we find unsatisfactory and do not support some of its claims, see details below.
However, the authors of all of these works, even though starting with rather general 2HDM,
quickly focus on some rather simple cases, which could be analyzed by simple algebra.
Thus, the most general 2HDM at finite temperature has never been analyzed up to now in any approximation.
\\

The structure of the paper is the following.
In Section 2 we briefly review the geometric approach to the most general 2HDM.
Then, in Section 3 we describe the phase diagram of the scalar part of the model.
In Section 4 we analyze thermal evolution of the ground state and
list possible sequences of phase transitions.
In Section 5 we discuss how symmetries of 2HDM are related with
possible thermal phase transitions.
Discussion and conclusions are presented in Sections 6 and 7.

%%%%%%%%%%%%%%%%%%%%%%%%%%%%%%%%%%%%%%%%%%%%%%
\section{Brief overview of the formalism}
%%%%%%%%%%%%%%%%%%%%%%%%%%%%%%%%%%%%%%%%%%%%%%%%

In this work we focus on the scalar sector of 2HDM.
The Higgs potential of the most general renormalizable 2HDM, $V_H = V_2 + V_4$, is
conventionally parametrized as
\bea
V_2&=&-{1\over 2}\left[m_{11}^2(\phi_1^\dagger\phi_1) +
m_{22}^2(\phi_2^\dagger\phi_2)
+ m_{12}^2 (\phi_1^\dagger\phi_2) + m_{12}^{2\ *} (\phi_2^\dagger\phi_1)\right]\,;\nonumber\\
V_4&=&\fr{\lambda_1}{2}(\phi_1^\dagger\phi_1)^2
+\fr{\lambda_2}{2}(\phi_2^\dagger\phi_2)^2
+\lambda_3(\phi_1^\dagger\phi_1) (\phi_2^\dagger\phi_2)
+\lambda_4(\phi_1^\dagger\phi_2) (\phi_2^\dagger\phi_1) \label{potential}\\
&+&\fr{1}{2}\left[\lambda_5(\phi_1^\dagger\phi_2)^2+
\lambda_5^*(\phi_2^\dagger\phi_1)^2\right]
+\left\{\left[\lambda_6(\phi_1^\dagger\phi_1)+\lambda_7
(\phi_2^\dagger\phi_2)\right](\phi_1^\dagger\phi_2) +{\rm
h.c.}\right\}\,.\nonumber
\eea
It contains 14 free parameters:
real $m_{11}^2, m_{22}^2, \lambda_1, \lambda_2, \lambda_3,
\lambda_4$ and complex $m_{12}^2, \lambda_5, \lambda_6,
\lambda_7$.
As a result the 2HDM a very rich phenomenology even at tree level.
However, different points in this 14-dimensional parameter space
do not necessarily correspond to distinct physics.
Indeed, if we perform any linear transformation between doublets
$\phi_1$ and $\phi_2$, we arrive at the same generic potential with
redefined coefficients, which still corresponds to the same set of observables.
Thus, the problem of minimizing the Higgs potential
has a reparametrization invariance, with reparametrization group $GL(2,C)$.
This property is an extension of an earlier identification of
rephasing $U(1)$, and unitary reparametrization $U(2)$, invariance of the model,
\cite{basis-inv,nishi,GK05,repfreedom};
it was developed in detail in \cite{Ivanov1,Ivanov2} and helped
prove several properties of the general 2HDM.
In this Section we will briefly review this formalism.

It starts with the observation of the internal Minkowski-space structure behind 2HDM.
We first introduce the four-vector
$r^\mu = (r_0,\,r_i) = (\Phi^\dagger \Phi,\, \Phi^\dagger \sigma^i \Phi)$,
where $\Phi = (\phi_1,\,\phi_2)^T$ is a 2-dimensional vector of Higgs doublets
and $\sigma^i$ are the Pauli matrices. This four vector is gauge invariant
and parametrizes the gauge orbits in the space of the Higgs fields.

The general reparametrization group $GL(2,C)$ can be written as $\mathbb{C}^*\times SL(2,C)$,
where $\mathbb{C}^*$ is the group of simultaneous multiplication of both $\phi_i$ with the same complex number,
while $SL(2,C)$ is the special linear transformation group containing all unit-determinant transformation matrices.
Multiplication of both doublets by the same number gives a freedom to arbitrarily rescale $r^\mu$,
however it does not modify the structure of the Higgs potential.
The special linear group, on the contrary, leads to non-trivial transformations of the Higgs potential,
and in subsequent analysis we focus on it.

Transformations of $\Phi$ under the special linear group $SL(2,C)$ corresponds
to the $SO(1,3)$ transformations of $r^\mu$, equipping the gauge orbit space with
the Minkowski-space structure.
It follows from the definition of $r^\mu$ that
\bea
&& r_0 = (\phi_1^\dagger\phi_1)+ (\phi_2^\dagger\phi_2) \ge 0\,,\nonumber\\
&& r^\mu r_\mu = 4\left[(\phi_1^\dagger\phi_1)(\phi_2^\dagger\phi_2)- (\phi_1^\dagger\phi_2)(\phi_2^\dagger\phi_1)\right]\ge 0\,. \label{from-definition}
\eea
Note that $r^\mu r_\mu >0$ is possible thanks to the fact that $\phi_1$ and $\phi_2$ are themselves
electroweak doublets, and the last line of (\ref{from-definition}) is just the Schwarz inequality written for $\phi_i$. 
Thus, the physically realizable vectors $r^\mu$
populate not the entire $1+3$-dimensional Minkowski space, but the future lightcone.
The Higgs potential in the $r^\mu$-space can be written in a very compact form:
\be
V = - M_\mu r^\mu + {1\over 2}\Lambda_{\mu\nu} r^\mu r^\nu\,.\label{Vmunu}
\ee
Here the four-vector $M_\mu$ is built from parameters $m_{ij}^2$ in (\ref{potential}),
while the symmetric four-tensor $\Lambda_{\mu\nu}$ is constructed from
the quartic coefficients $\lambda_i$.
Their explicit expressions as well as some properties can be found in \cite{Ivanov1,Ivanov2}.
Here we just note the most important property of $\Lambda_{\mu\nu}$:
it can always be diagonalized by a specific $SO(1,3)$ transformation of the $r^\mu$-space,
and after diagonalization it takes form
\be
\Lambda_{\mu\nu} = \mathrm{diag}(\Lambda_0,\,-\Lambda_1,\,-\Lambda_2,\,-\Lambda_3)\quad
\mathrm{with} \quad \Lambda_0>0,\quad \Lambda_0>\Lambda_i\,,\ i=1,2,3\,,\label{Lambdamunudiag}
\ee
where the inequalities among the eigenvalues result from the positivity constraint on the potential.
The minus signs in front of the ``space-like'' eigenvalues arise from the pseudo-euclidean
metric in the orbits space.

It is known that the potential (\ref{Vmunu}) can have three types of minima:
(i) the electroweak (EW) conserving,
(ii) the EW-breaking but charge conserving (i.e. neutral), and
(iii) the EW- and charge-breaking ones.
One can use the v.e.v.s of the two doublets $\lr{\phi_i}$ to construct $\lr{r^\mu}$.
Then, the three type of minima correspond to:
(i) $\lr{r^\mu} = 0$ (the apex of the forward lightcone $LC^+$),
(ii) $\lr{r^\mu} \not = 0$ but $\lr{r^\mu}\lr{r_\mu} = 0$ (the surface of $LC^+$),
(iii) $\lr{r^\mu} \not = 0$ and $\lr{r^\mu}\lr{r_\mu} > 0$ (the interior of $LC^+$).

The position of the charge-breaking extremum $\lr{r_\nu}_{ch}$ is given by the following
coupled equations:
\be
\Lambda^{\mu\nu} \lr{r_\nu}_{ch} = M^\mu\,,\label{extremum1}
\ee
If $\Lambda^{\mu\nu}$ is not singular, a solution of this system always exists and is unique.
However solutions realizable in the 2HDM, i.e. such that $\lr{r_\nu}_{ch}$ lies inside the forward lightcone,
exist only if $M^\mu$ lies inside some cone with the apex at the origin.
If this is the case, then an additional requirement ($\Lambda_i < 0$, $i=1,2,3$,
i.e. the tensor $\Lambda_{\mu\nu}$ is positive-definite in the entire space of non-zero vectors $r^\mu$)
guarantees that the charge-breaking extremum is a minimum, \cite{Ivanov1}. If, on the contrary, 
at least one of $\Lambda_i$ is positive, then the charge-breaking extremum is a saddle point.

The positions of all neutral extrema $\lr{r^\mu}$ are the solutions of the following coupled equations:
\be
\Lambda^{\mu\nu} \lr{r_\nu} - \zeta \lr{r^\mu} = M^\mu\,,\quad \lr{r^\mu}\lr{r_\mu} = 0\,,\label{extremum2}
\ee
where $\zeta$ is a Lagrange multiplier. This system can have up to six solutions.
To distunguish a (local) minimum from other types of extrema one needs to use more involved tools,
for example, geometric methods developed in \cite{Ivanov1,Ivanov2}.
These methods allowed to prove that the generic potential has at most two local minima
in the entire orbit space (not just on the surface of the cone), while the other extrema are saddle points.

Finally, the Higgs potential (\ref{potential})
must be accompanied by the kinetic term, which can also be written covariantly:
\be
K = K_\mu \rho^\mu\,,\quad \rho^\mu = (D_\alpha \Phi)^\dagger \sigma^\mu (D^\alpha \Phi)\,,
\ee
where $D_\alpha$ is the covariant derivative, $\alpha$ denotes the usual space-time coordinates,
while $\mu$, as before, denotes the coordinate in the orbit space.
Note that reparametrization transformation properties of $\rho^\mu$ are the same as $r^\mu$.
In the default frame, $K^\mu = (1,\,0,\,0,\,0)$. Upon an $SO(1,3)$ transformation, $K^\mu$
acquires non-zero ``space-like'' coordinates, however the condition $K^\mu K_\mu = 1$ is always satisfied
(if we considered $GL(2,C)$ transformations of the fields instead of $SL(2,C)$, this would not hold anymore;
however we do not use the rescaling subgroup of $GL(2,C)$ just because 
the $SL(2,C)$ group itself is sufficient in order to describe the phases 
and the sequences of possible phase transitions of 2HDM).
The four-vector $K^\mu$ is not involved in the search for the minimum of the potential,
however it affects the mass matrix at this minimum.

This generalized kinetic term effectively incorporates the non-diagonal kinetic term, which, as was
argued in \cite{ginz-kinetic}, must be introduced in the initial lagrangian to restore renormalizability of the model.

\section{The phase diagram of the scalar sector of 2HDM}

Before studying thermal evolution of the 2HDM vacuum, let us explicitly describe the phase diagram
of the zero-temperature 2HDM.
Here we use results from \cite{Ivanov2,GL} but slightly reformulate
them in anticipation of the study of thermal evolution.

Note that whenever we say ``phase transition'' in this Section, we imply
zero-temperature phase transition, that is, transitions from one phase to another 
upon continous change of the parameters of the potential.
We will classify such zero-temperature phase transitions as 
the ``second-order'' and ``first-order'' ones according to whether
the order parameter (the position of the global minimum of the potential) changes
continuously or not. 

We first switch to the $\Lambda_{\mu\nu}$-diagonal frame.
We will focus here on the generic situation, in which
all the ``space-like'' eigenvalues $\Lambda_i$ are different.
The phases can be then represented as regions in the $M^\mu$-space in this frame.

The structure of the phase diagram depends on the sign of $M_0$.

\subsection{The lower half-space}

In the lower half-space of $M^\mu$, i.e. with $M_0\le 0$,
the phase diagram is extremely simple.
\begin{itemize}
\item
If $M^\mu M_\mu \ge 0$, the only extremum of the potential is the minimum at the origin.
Thus, the past lightcone in the $M^\mu$-space is the EW-symmetric phase.
\item
If $M^\mu M_\mu < 0$, i.e. $M^\mu$ lies outside the past lightcone but still in the lower half-space,
then there exists a unique non-zero neutral extremum, which is necessarily the global minimum.
\end{itemize}
Thus, in the lower half-plane we have only one surface of phase transitions, the
past lightcone, at which the EW-breaking/restoring phase transition takes place.

\subsection{The upper half-space}

For $M_0>0$, the phase diagram is much richer. To describe it,
let us introduce the 3-vector $\mathfrak{m}_i$, $i=1,2,3$:
\be
\mathfrak{m}_i = \dfrac{1}{M_0} (M_1,\,M_2,\, M_3)\label{mfrak}
\ee
and show the phase diagram in the $\mathfrak{m}_i$-space.

\begin{figure}[htb]
\begin{center}
\includegraphics[width=0.4\textwidth]{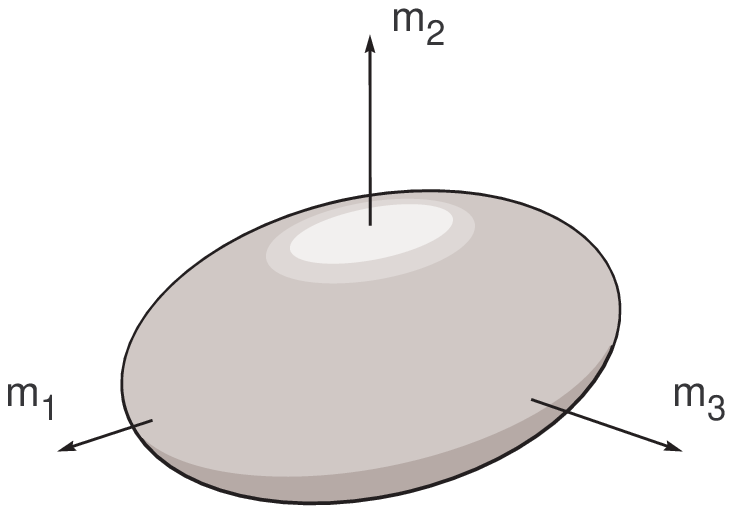}
\includegraphics[width=0.4\textwidth]{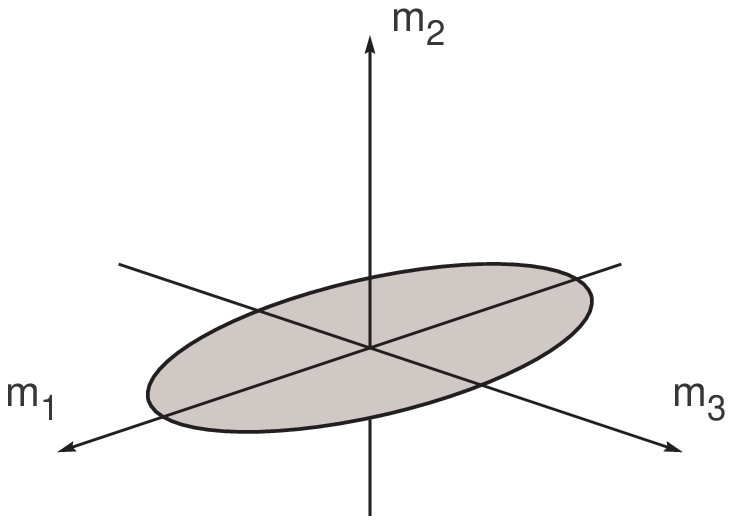}
\caption{The phase diagram in the $\mathfrak{m}_i$-space.
{\em Left}: all $\Lambda_i <0$; the ellipsoid separates the charge-breaking
and the neutral vacua. {\em Right}: $\Lambda_2$ is positive
and greater than $\Lambda_{1}$ and $\Lambda_{3}$. Inside the dark ellipse 
the discrete symmetry of the potential $V$ is spontaneously broken.}
  \label{fig-3D-phase}
\end{center}
\end{figure}

There are two generic cases to consider, which are shown in Fig.~\ref{fig-3D-phase}.

\subsubsection{All $\Lambda_i<0$}

The condition $\Lambda_1,\,\Lambda_2,\,\Lambda_3 < 0$ is the necessary condition for the charge-violating minimum to exist.
Whether this minimum is indeed realized, depends on $\mathfrak{m}_i$, i.e. on the point of the phase diagram.
Indeed, in the $\Lambda_{\mu\nu}$-diagonal frame, Eq.~(\ref{extremum1}) can be rewritten as
\be
\Lambda_0 \lr{r_0} = M_0\,,\quad \Lambda_i \lr{r_i} = M_i\,.
\ee
Since the solution $\lr{r^\mu}$ must lie inside the forward lightcone, one obtains that
the charge-breaking phase in the phase diagrams lies inside the ellipsoid
\be
{\mathfrak{m}_1^2 \over a_1^2} + {\mathfrak{m}_3^2 \over a_3^2} + {\mathfrak{m}_3^2 \over a_3^2}< 1\,,\quad
\mathrm{where}\quad
a_i = {|\Lambda_i| \over \Lambda_0}\,,\quad i=1,2,3\,,\label{ellipsoid}
\ee
shown in Fig.~\ref{fig-3D-phase}, left.

If $\mathfrak{m}_i$ lies outside ellipsoid (\ref{ellipsoid}), then the vacuum is neutral.
The surface of the ellipsoid is the locus of the critical points of the phase diagram,
at which the second order charge-breaking or charge-restoring phase transition takes place.

\subsubsection{At least one $\Lambda_i >0$}

If at least one $\Lambda_i$ is positive, then the minimum always corresponds to a neutral vacuum 
and its position satisfies equations (\ref{extremum2}).
Writing $\lr{r^\mu}$ as $r_0(1,\,n_1,\,n_2,\,n_3)$ with $|\vec n|=1$,
one can eliminate the Lagrange multiplier from (\ref{extremum2}) to arrive
at the following system of equations:
\be
[M_0 - (\Lambda_0-\Lambda_i)r_0]n_i = M_i\,,\quad i=1,2,3\,.\label{extremum3}
\ee
The requirement that $\vec n$ is a unit vector can be written as
\be
\sum_i { \mathfrak{m}_i^2 \over \left(1 - r_0 {\Lambda_0-\Lambda_i \over M_0}\right)^2} = 1\,.\label{sixth}
\ee
This is a sixth order equation for $r_0$.

In a general case, the system cannot be solved exactly.
However, the geometric approach of \cite{Ivanov1,Ivanov2}
still allows one to understand the structure of the phase diagram.

First, if $\mathfrak{m}_i$ lies on one of the principal planes, then the potential
has an additional discrete symmetry.
The solutions of (\ref{extremum3}) can either conserve or violate this symmetry.
The necessary and sufficient condition for violation of this symmetry, e.g. in the case
$\mathfrak{m}_2=0$, are
\be
{\mathfrak{m}_1^2 \over b_1^2} + {\mathfrak{m}_3^2 \over b_3^2} < 1\,,\quad
\mathrm{where}\quad
b_i = {\Lambda_2 - \Lambda_i \over \Lambda_0 - \Lambda_2}\,,\quad i=1,3.
\ee
Thus, symmetry-violating extrema appear, if $\mathfrak{m}_i$ lies inside a specific ellipse on one of the
principal planes.

In \cite{Ivanov1,GL} it was proved that these symmetry-violating extrema are minima, if and only if
the corresponding eigenvalue of $\Lambda_{\mu\nu}$ ($\Lambda_2$ for $\mathfrak{m}_2=0$) 
is positive and is the biggest of all $\Lambda_i$.
So, when constructing the phase diagram, we should first identify the largest eigenvalue
and then consider only the ellipse that lies in the plane orthogonal to the corresponding eigenvector,
as it is shown in Fig.~\ref{fig-3D-phase}, right.
For points lying strictly inside the ellipse two different degenerate minima exist.
In both of them the discrete symmetry is spontaneously broken\footnote{With an appropriately symmetric
kinetic term $K^\mu$, this extends to the symmetry of the whole scalar lagrangian.
The case depicted in Fig.~\ref{fig-3D-phase}, right, corresponds then to spontaneous $CP$-violation.}.
For points in the $(\mathfrak{m}_1,\mathfrak{m}_3)$ plane outside the ellipse only one minimum exists
and the discrete symmetry is preserved.
So, if $\mathfrak{m}_i$ lies strictly on the plane and moves from outside into the ellipse,
a symmetry-breaking second-order phase transition takes place.

The points just above ($\mathfrak{m}_2>0$) or just below ($\mathfrak{m}_2<0$) the ellipse have two minima at different depths.
So, if $\mathfrak{m}_i$ lies above the ellipse and moves through it downwards,
the relative depth between the two minima changes sign, and a first-order phase transition
takes place\footnote{We consider here only the phase diagram at equilibrium and do not discuss
the actual kinetics of this phase transition.}.
Thus, the {\em interior} of the ellipse is the locus of the first-order phase transitions,
while its {\em boundary} is the locus of the second-order phase transitions.

Let us also discuss the phase diagram of 2HDM obtained in \cite{Gufan}.
The authors of that paper used not the most general, but a rather restricted version of the 2HDM potential,
with only 8 non-zero free parameters out of 14. The parameters $\lambda_6$, $\lambda_7$ and $m_{12}^2$ were set to zero,
so that the potential from the start was $(Z_2)^2$-symmetric.
This choice authomatically excludes any $CP$-violation, which is the origin of the authors' surprising
assertion that the relative phase between the two v.e.v.'s ``is not demonstrated in the particle properties''.

The potential chosen in \cite{Gufan} was sufficiently simple to allow for
straightforward minimization. Six phases were identified and classified according to the residual symmetry and
the ``number of condensates'', i.e. whether $\lr{\fd_1\f_1}$ and $\lr{\fd_1\f_2}$ are non-zero.
We find this picture misleading for the following reason. If one focuses on the scalar part of the theory only,
then all the observables, including classification of the phases according to the properties of the ground state,
must remain invariant under the group of the most general reparametrization transformations of the scalar lagrangian.
However, the classification of \cite{Gufan} is not reparametrization invariant.
Phases II, III and IV are all related by reparametrization transformations and correspond
just to the neutral vacuum, while phases V and VI correspond to the charge-breaking vacuum.
This is not surprising since the form of the 2HDM potential used in \cite{Gufan} was rather restrictive
and obscured these relations.
Therefore, some of their phases do indeed correspond to truly different 2HDM phases:
the governing principle is not the number of condensates, but the symmetries of the lagrangian
and of the ground state.

%%%%%%%%%%%%%%%%%%%%%%%%%%%%%%%%%%%%%%%%%%%%%%
\section{Thermal evolution of the ground state}
%%%%%%%%%%%%%%%%%%%%%%%%%%%%%%%%%%%%%%%%%%%%%%%%

\subsection{Temperature dependence of the mass terms}

Let us now consider the thermal evolution of the ground state of the model.
As said in the Introduction, our motivation is not to provide a phenomenologically accurate description
of thermal 2HDM, but rather to extend the developed formalism, which effectively works for the most general
2HDM, to non-zero temperatures. To this end, we stick to the following approximations:
\begin{itemize}
\item
we consider only the scalar sector of 2HDM
and neglect temperature corrections from gauge bosons and the fermions,
\item
we consider only the largest one-loop high-temperature corrections $\propto \lambda_i T^2$,
\item
although the results can be trusted only at temperatures much larger than the mass parameters, 
we still use them to discuss the lower temperature region, $T\sim m$.
\end{itemize}
Although in this region our results must be strongly corrected by terms which we neglect,
we still think our discussion provided a first useful glimpse at the typical patterns
of phase transitions one can expect in 2HDM.
It is possible that the approach used here can be developed further to remove some of these assumptions.

The largest thermal correction to the scalar $\phi^4$-theory comes from the hard thermal
one-loop self-energy diagrams, see e.g. \cite{kapusta}.
In the one-component real scalar field theory
with negligible zero-temperature mass and interaction term $\lambda \phi^4/4!$,
the thermal correction is
$$
\delta m^2 = {\lambda T^2 \over 24}\,.
$$
In 2HDM, we have similar diagrams with various fields in the loop.
In order to extract the largest, $\propto \lambda_i T^2$ correction, we neglect mass terms
in the propagators in loops, so that the scalar loops generate only a universal factor.
Thus, the temperature corrections to $m_{ij}^2$ differ only in vertex factors,
and the straightforward calculation gives:
\bea
m_{11}^2(T) & = & m_{11}^2(0)- T^2 \cdot {3\lambda_1+2\lambda_3+\lambda_4 \over 6}\,,\nonumber\\[2mm]
m_{22}^2(T) & = & m_{22}^2(0)- T^2 \cdot {3\lambda_2+2\lambda_3+\lambda_4 \over 6}\,,\nonumber\\[2mm]
m_{12}^2(T) & = & m_{12}^2(0)- T^2 \cdot {\lambda_6+\lambda_7 \over 2}\,. \label{Tempevolgen}
\eea
As we already said, strictly speaking, these equations are reliable only at temperatures much larger then masses of the particles
in loops. At lower temperatures, the mass parameters cannot be neglected anymore,
which significantly complicates the problem.
However, the present study is only the first approximation to a more complete analysis of thermal 2HDM,
so we adopt (\ref{Tempevolgen}) at all temperatures and follow its physical consequences.

In the Minkowski-space formalism, one can rewrite (\ref{Tempevolgen}) as the temperature dependence
of the $SO(1,3)$-covariant four-vector $M^\mu$.
Indeed, in the Appendix we show that
\be
M^\mu(T) = M^\mu(0) - c^\mu\cdot T^2\,,\qquad
c_\mu = {1 \over 12}\left(6 \Lambda_{\mu\nu} - g_{\mu\nu}\mathrm{Tr}\Lambda\right) K^\nu\,,\label{MmuT}
\ee
where $\mathrm{Tr}\Lambda \equiv \Lambda_{\mu\nu}g^{\mu\nu} =
\Lambda_0 + \Lambda_1 + \Lambda_2 + \Lambda_3 = \lambda_3 - \lambda_4$.
The four-vector of kinetic terms $K^\nu$ appears in this result
due to the presence of the loop propagator. Note that $K^\nu$ does not receive
any temperature corrections within this approximation.

Expression (\ref{MmuT}) can be used in any frame.
In particular, in the original frame it reproduces (\ref{Tempevolgen}),
while in the $\Lambda_{\mu\nu}$-diagonal frame it gives:
\bea
&& c_0 = {6\Lambda_0 - \mathrm{Tr}\Lambda \over 12}K_0 = {5\Lambda_0 - \Lambda_1 - \Lambda_2 - \Lambda_3 \over 12} K_0 > 0\,,\\[2mm]
&& c_i = {6\Lambda_i - \mathrm{Tr}\Lambda \over 12}K_i\,\quad i=1,2,3\,.
\eea
Note that $c_i$ can have any sign depending on the values of $\Lambda_i$.

One can ask whether there is any constraint on the possible values of $c_i$. The answer is in the affirmative.
Since $K^\mu K_\mu = 1$, one obtains from (\ref{MmuT}):
\be
\sum_i {c_i^2 \over c_0^2}\cdot \left({6\Lambda_0 - \mathrm{Tr}\Lambda \over 6\Lambda_i - \mathrm{Tr}\Lambda}\right)^2 =
\sum_i {K_i^2 \over K_0^2} < 1\,.
\ee
Thus, possible $c_i/c_0$ lie inside the ellipsoid, whose semiaxes are always smaller than 5.

Consider next $M_0(T)$ in this frame: $M_0(T) = M_0 - c_0 T^2$. If $M_0 <0$, it further decreases 
with the increasing temperature without changing its sign.
Thus, the only phase transition one can have in this case is the EW-restoring
phase transition.

If the initial $M_0>0$, it will change its sign at
\be
T = T_* \equiv \sqrt{{M_0 \over c_0}}\,.
\ee
Then one can separate thermal evolution of the ground state into two stages,
which can be dubbed the ``low-temperature'' ($T<T_*$) and the ``high-temperature''
($T>T_*$) stages.
Again, it is only the low-temperature stage that non-trivial phase transitions can take place,
while during the high-temperature stage one can observe only the EW-restoring phase transition.

Focusing on the low-temperature stage, let us define $\mathfrak{m}_i(T)$ similarly to (\ref{mfrak}):
\be
\mathfrak{m}_i(T) = \dfrac{M_i(T)}{M_0(T)} =
\left({M_1 - c_1 T^2 \over M_0 - c_0 T^2},\,
{M_2 - c_2 T^2 \over M_0 - c_0 T^2},\,
{M_3 - c_3 T^2 \over M_0 - c_0 T^2}\right)\,.
\label{mfrakT}
\ee
If we introduce $x = T^2/(T_*^2 - T^2)$, so that $x=0$ at zero temperature, then
\bea
\mathfrak{m}_i(T) = \mathfrak{m}_i \left[ 1+ \left(1- {c_i \over c_0 \mathfrak{m}_i}\right)x\right]
& \mathrm{if} & \mathfrak{m}_i \not = 0\,,\nonumber\\[2mm]
\mathfrak{m}_i(T) = - {c_i \over c_0}x\,,& \mathrm{if} &\mathfrak{m}_i = 0 \,.\label{mfrakTi}
\eea
Thus, the temperature evolution $\mathfrak{m}_i(T)$ is represented on the phase diagram by a {\em straight ray}
starting from $\mathfrak{m}_i$ at $T=0$ and going to infinity as $T \to T_*$.

It is easy to see that this result is not specific for the quadratic temperature dependence.
It is based solely on the fact that the functional form of the
temperature dependent correction to $m_{ij}^2$ is the same.

\subsection{Sequences of thermal phase transitions}

The fact that thermal evolution of $\mathfrak{m}_i$ is represented by a straight ray
allows us to immediately classify possible sequences of phase transitions without any
further calculation.

Depending on the eigenvalues $\Lambda_i$, the sign of each of the components $\mathfrak{m}_i(T)$ in (\ref{mfrakTi}) 
can either remain unchanged or can change only once at some temperature.

If the signs of all $\mathfrak{m}_i(T)$ do not change,
the only phase transition one can have at growing temperature
is the charge-restoring second-order phase transition,
which takes place for negative $\Lambda_i$ and $\mathfrak{m}_i$ inside the ellipsoid,
or the discrete symmetry restoring phase transition, when at least one $\Lambda_i$ is positive
and $\mathfrak{m}_i$ lies inside the corresponding planar ellipse.

\begin{figure}[htb]
\begin{center}
\includegraphics[width=0.3\textwidth]{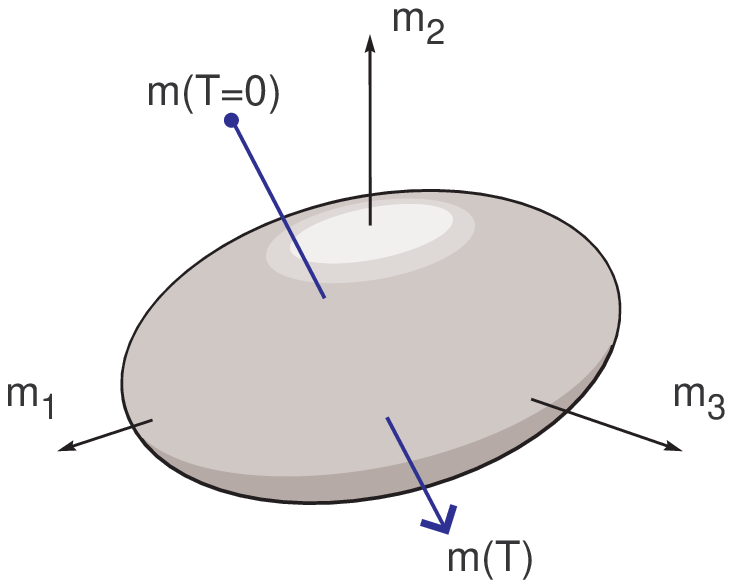}
\includegraphics[width=0.3\textwidth]{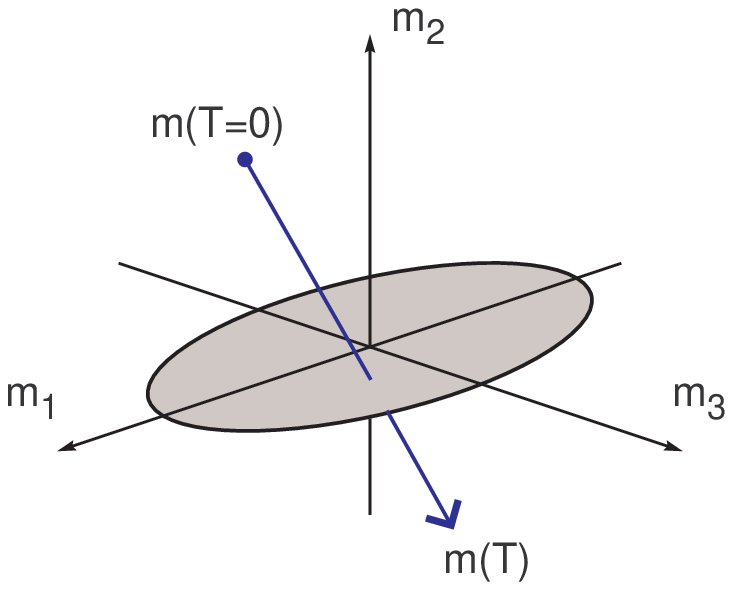}
\includegraphics[width=0.3\textwidth]{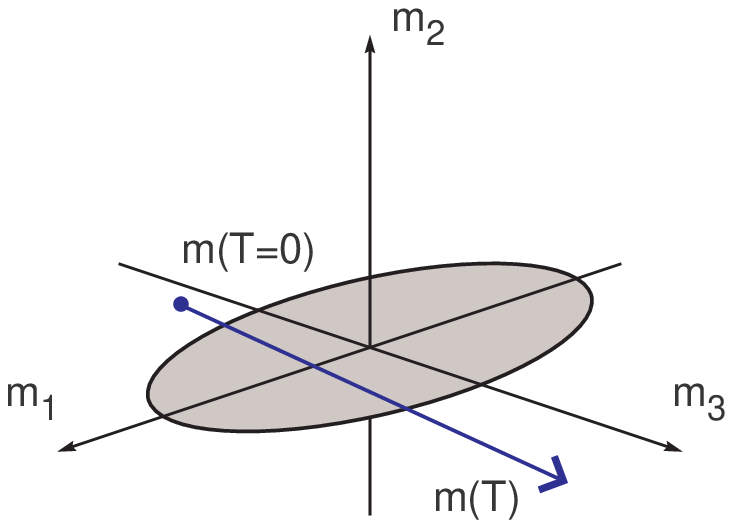}
\caption{Examples of the temperature evolution trajectories (blue rays)
for the two phase diagrams shown in Fig.~\ref{fig-3D-phase}. In the rightmost figure the ray lies
in the plane of the ellipse.}
  \label{fig-3D-path}
\end{center}
\end{figure}

If one or more $\mathfrak{m}_i(T)$ change their signs, then one can have a richer spectrum of possibilities.
On its way to infinity, the ray can go through various intermediate phases.
In Fig.~\ref{fig-3D-path} we show examples of such paths for the two phase diagrams
shown previously in Fig.~\ref{fig-3D-phase}.
In Fig.~\ref{fig-3D-path}, left, we show that the charge-breaking phase can be an intermediate phase
of thermal evolution of the model. Fig.~\ref{fig-3D-path}, middle, shows
a path that leads to a first-order phase transition, while Fig.~\ref{fig-3D-path}, right, demonstrates
a planar ray that goes exactly through the ellipse and leads to a sequence of two
second-order phase transitions: the symmetry-breaking and the symmetry-restoring ones.

By combining various initial positions and directions of the thermal evolutions,
one can easily generate a list of various sequences of phase transitions.
However, from general geometric considerations (a straight line can intersect an ellipse or ellipsoid
no more than twice), it follows that the longest chain of phase transitions consists
of two phase transitions (followed by the electroweak phase transition at very high temperatures).
Examples are shown in Fig.~\ref{fig-3D-path}.

It can also happen that the ray barely touches but does not intersect the ellipsoid
of Fig.~\ref{fig-3D-phase}, left, or the ellipse of Fig.~\ref{fig-3D-phase}, right.
In such cases, the temperature evolution drives the system to a critical point and then returns back
to the initial phase.

\section{Symmetries}

The scalar sector of the two-Higgs-doublet model can have various additional explicit symmetries
\cite{Ivanov2,GK05}.
Suppose that 2HDM possesses some explicit symmetry at zero temperature. One can then ask:
\begin{itemize}
\item
does thermal evolution preserve this symmetry?
\item
how are the possible sequences of phase transitions restricted by such a symmetry?
\end{itemize}

A detailed analysis of possible symmetries of 2HDM was carried out in \cite{Ivanov2}.
It was stressed there that one should distinguish between symmetries of the potential and
of the entire lagrangian, since they play different roles in the theory.
A symmetry of the potential governs the vacuum structure and possible degeneracies,
while a symmetry of the full Higgs lagrangian is needed to make the mass matrix symmetric, too.

Symmetries are best explored in the $\Lambda_{\mu\nu}$-diagonal frame.
If all $\Lambda_i$ are different, then in this frame the quartic part of the potential has symmetry group $(Z_2)^3$ generated by
reflections of the three directions. If some of the spacelike components of $M^\mu$ are zero in this frame,
then the potential has some non-trivial explicit symmetry. If, in addition, the corresponding components
of $K^\mu$ are also zero, then this symmetry extends to the entire lagrangian.

Consider now Eq.~(\ref{MmuT}), which gives $M^\mu(T)$.
It is easy to see that if at $T=0$ there is a symmetry
of the entire lagrangian (i.e. of $\Lambda_{\mu\nu}$, $M^\mu$ and $K^\mu$),
then it will be preserved during the entire thermal evolution.
If, instead, only the Higgs potential, but not the entire lagrangian, has some symmetry,
then {\em generically} this symmetry is explicitly violated at non-zero temperature\footnote{One must of course
keep in mind that a symmetry of the potential not shared by the kinetic term is not stable in the perturbative series. 
A discussion involving such a symmetry is thus restricted to the tree-level approximation of the potential.}.
There is, however, an exceptional situation when $\Lambda_i$ satisfy the relation $\mathrm{Tr}\Lambda = 6 \Lambda_i$,
so that the corresponding correction coefficient $c_i = 0$. In this case, the symmetry
of the potential is preserved during the temperature evolution.

Let us now take a closer look at the first-order phase transition, which happens when the
evolution path crosses the ellipse of Fig.~\ref{fig-3D-path}, middle.
Exactly at the phase transition there are two separate degenerate minima of the Higgs potential.
From Proposition~4 of \cite{Ivanov2} it follows that at this point (and only at this point!)
a discrete symmetry of the potential emerges. At higher temperature this symmetry is lost again,
therefore, it cannot be the symmetry of the entire lagrangian. Thus, we arrive at the following conclusion:\\

\noindent{\em The first-order phase transitions are associated with points of a discrete symmetry of the potential,
but not of the entire lagrangian}.\\

One concludes that the symmetries which are often studied in literature, like $CP$
or the so-called $Z_2$-symmetry (which is in fact a $(Z_2)^2$-symmetry, see \cite{Ivanov2}) cannot be associated with
first-order phase transitions in 2HDM.
An example of symmetry that can be associated with the first-order phase transition will be given
in \cite{GIKprep}.

Of course, this result is derived only within the approximation chosen. Higher-order corrections to the free energy potential
can turn some of the second-order phase transitions into the first-order ones.

Let us also discuss the phase transitions identified in \cite{Gufan}.
Similarly to our analysis, the authors of that paper assumed that the thermal evolution can still be described
by the same fourth-order free energy density, but with the coefficients depending on temperature.
They, however, did not calculate this dependence, thus overlooking a possibility of electroweak symmetry non-restoration
at large temperature. Among the other results, they find the possibility of first-order phase transition
between the charge-breaking and the neutral vacua (in their notation, between phases IV and V).
Our analysis does not leave any room for such a phase transition.
The first-order phase transitions can arise only when two separate local minima coexist.
However, we proved in \cite{Ivanov1} that disjoint charge-breaking and neutral minima never coexist
in any 2HDM potential. Therefore, charge-breaking and charge-restoring transitions occur
via a continuous change of the vacuum expectation values. The depth of the global minimum then changes smoothly,
and there is no specific heat associated with such a phase transition.

In fact, within the approximation chosen a first order phase transition can only take place between the same symmetry phases
(isostructural phase transition).

%%%%%%%%%%%%%%%%%%%%%%%%%%%%%%%%%%%%%%%%%%%%%%
\section{Discussion}
%%%%%%%%%%%%%%%%%%%%%%%%%%%%%%%%%%%%%%%%%%%%%%%%

\subsection{Non-restoration of the EW symmetry}

Since the seminal paper \cite{Weinberg} of Weinberg it is known that a symmetry, global or local, 
can remain broken even at very high temperatures.
The simplest example of such a case is given by a $\phi^4$-model with two real scalars, \cite{Dvali95}.
This possibility is of course present in 2HDM, as can be seen directly from eq.(\ref{Tempevolgen}) 
giving the temperature evolution of the mass terms $m_{ij}^2(T)$.
This happens, for example, when $\lambda_6=\lambda_7=0$, and
$3\lambda_1 + 2\lambda_3 + \lambda_4$ and $3\lambda_2 + 2\lambda_3 + \lambda_4$ are of opposite signs.
Together with the positivity conditions this typically implies
that $\lambda_1 \gg \lambda_2$ (or vice versa), with $2\lambda_3 + \lambda_4$ sufficiently negative.

In the Minkowskian formalism, this situation takes place, when $M^\mu(T)$ goes down in the phase diagram,
but not sufficiently steeply to cross the past lightcone,
i.e. if the four-vector $c^\mu$ defined in (\ref{MmuT}) is spacelike, $c^\mu c_\mu <0$:
\be
36 (\Lambda^2)_{\mu\nu} K^\mu K^\nu - 12 \mathrm{Tr}\Lambda\cdot \Lambda_{\mu\nu} K^\mu K^\nu + (\mathrm{Tr}\Lambda)^2 < 0\,.
\ee
A necessary, but not sufficient, condition for this to take place is that there exists at least one negative $\Lambda_i$.

Note also that the electroweak symmetry is the only symmetry that might remain
broken at very large temperatures.
All other symmetries are unaviodably restored, as it is obvious from the analysis of the phase diagram.
This generalizes the result obtained in \cite{Dvali96} that the spontaneously broken $CP$-symmetry
is always restored at high temperatures in the 2HDM.

\subsection{Beyond the approximation chosen}

Within the approximation chosen, the above discussion gives an exhaustive list of sequences
of phase transitions possible in 2HDM. However, various additional corrections
will definitely modify this picture.

Firstly, for temperatures $T$ comparable with the mass parameters the loop integrals determining 
the corrections to the mass terms of the potential are not universal. As a result, the functional 
$T$-dependence of these terms is more complicated and the thermal evolution will not follow 
the straight rays on the phase diagram.
Second, there will be temperature corrections to $\lambda_i$, which
will result in temperature evolution of the phase diagram itself.
Third, the other fields present in 2HDM will contribute to the effective potential and will
affect evolution of the ground state.

It remains to be seen whether the formalism used here can be extended further to cope with these terms,
still preserving the generality of the model.

\section{Conclusions}

The two-Higgs-doublet model has a rich phase diagram even
in the tree-level approximation. One can expect a similarly rich
spectrum of phase transitions of 2HDM at non-zero temperatures.
Unfortunately, analysis of thermal evolution of the most general 2HDM has been impossible
for a long time due to algebraic complexity of the problem.
In this paper, armed with the recently developed geometric approach to the most general
renormalizable 2HDM, we get an insight into various properties of the model
at non-zero temperature.

We have computed the largest temperature corrections to the mass terms in a basis-invariant way and
demostrated that, in the approximation adopted, the thermal evolution of the ground state
is described by a straight ray on the fixed phase diagram.
This allowed to establish all possible sequences of phase transitions within this approximation.
Among the results that we have obtained are: the possibility to have spontaneously $CP$-violating
or charge-violating ground states as {\em intermediate} stages of the thermal evolution,
and the possibility to have first order phase transitions associated with a momentary restoration
of a discrete symmetry of the potential, but not of the entire scalar lagrangian.

These findings might have intriguing cosmological consequences, which will be studied
in a more phenomenology-oriented paper \cite{GIKprep},
see also some preliminary discussion in \cite{Gin06}.
The purpose of this work was just to show how to use the geometric approach to 2HDM to get insight
into its thermal dynamics.
\\

{\bf Acknowledgements.}
I am thankful to Ilya Ginzburg and Kostya Kanishev for discussions and useful comments, as well as
to the referee for numerious suggestions for improvement of the text.
This work was supported by the Belgian Fund F.R.S.-FNRS via the
contract of Charg\'e de recherches and in part by grants
RFBR 08-02-00334-a and NSh-1027.2008.2

\appendix

\section{Derivation of the covariant temperature corrections}

Here we derive formula (\ref{MmuT}) for the temperature corrections to the mass terms
in the covariant form.

We start with the Higgs lagrangian in the tensorial form:
\be
L = {K^a}_b (\partial_\mu\f^*_a) (\partial^\mu\f_b) +
{Y^a}_b (\f^*_a \f_b) - {{{Z^a}_b}^c}_d (\f^*_a \f_b)(\f^*_c \f_d)\,. \label{temperature.1}
\ee
Note also that to make the expression more symmetric,
we assumed in (\ref{temperature.1}) that all terms are written
{\em separately}, i.e. instead of $\lambda_3(\fd_1 \f_1)(\fd_2 \f_2)$ we write
\be
\dots + {1 \over 2}\lambda_3 (\fd_1 \f_1)(\fd_2 \f_2)
+ {1 \over 2}\lambda_3 (\fd_2 \f_2)(\fd_1 \f_1) + \dots
\ee
In contrast to the standard tensorial form, here indices $a\dots d$ are double indices:
$$
a = \{\alpha,\,i\}\,,\quad \alpha=1,2\,,\quad i = \uparrow,\downarrow\,.
$$
The subindex $\alpha$ labels the doublets and $i$ labels components of the doublet $\alpha$. The values of ${Y^a}_b$
and ${{{Z^a}_b}^c}_d$ are insensitive to $i$:
\bea
&&
{K^a}_b \equiv {K^{\{\alpha, i\}}}_{\{\beta, j\}} = K^\alpha{}_\beta\cdot \delta^i_j\,,\qquad
{Y^a}_b \equiv {Y^{\{\alpha, i\}}}_{\{\beta, j\}} = Y^\alpha{}_\beta\cdot \delta^i_j\,,\nonumber\\
&& \zz{a}{b}{c}{d} \equiv \zzz{\{\alpha, i\}}{\{\beta, j\}}{\{\gamma, k\}}{\{\delta, m\}}
= \zz{\alpha}{\beta}{\gamma}{\delta}\cdot \delta^i_j\delta^k_m\,.\label{temperature.2}
\eea
Then, the 1-loop correction to the self-energy is given by the contraction of the quartic coupling
with the propagator, which brings in $K^{-1}$:
\be
\delta {Y^a}_b  = - C \cdot \left({{{Z^a}_b}^c}_d + {{{Z^a}_d}^c}_b +
{{{Z^c}_b}^a}_d + {{{Z^c}_d}^a}_b\right) {(K^{-1})_c}^d
= - 2 C \left({{{Z^a}_b}^c}_d + {{{Z^a}_d}^c}_b \right) {(K^{-1})_c}^d \,,\label{temperature.3}
\ee
where $C = T^2/12$. Note that summation over all possible ways of contracting indices automatically takes into account
configurations with identical fields. One can then sum over intra-doublet indices $i$, which gives:
\be
\delta Y^\alpha {}_\beta = -2 C \left(2\zz{\alpha}{\beta}{\gamma}{\delta}+\zz{\alpha}{\delta}{\gamma}{\beta}\right)
{(K^{-1})_\gamma}^\delta\,.\label{temperature.4}
\ee
The extra factor 2 in the first term inside the brackets comes from $\delta_i^j \delta_i^j = 2$.
This expression is well-known, see e.g. Eq.~(11) in \cite{Dvali96}.

The next step is to translate (\ref{temperature.4}) to the covariant notation.
If $A$ is a Hermitian $2\times 2$ matrix, then one can write
$$
A = a_0 \sigma^0  - a^i \sigma^i \equiv a^\mu \sigma_\mu\,,
$$
where $a^\mu = (a^0,\,a^i)$ and $\sigma^\mu = (\sigma^0,\, \sigma^i)$. The inverse matrix can be
written using $\bar\sigma^\mu = (\sigma^0,\,-\sigma^i)$:
$$
A^{-1} = {a^\mu \bar\sigma_\mu \over a^2}\,, \quad a^2 = a^\mu a_\mu\,.
$$
If follows from definitions that $\zz{\alpha}{\beta}{\gamma}{\delta}$ can be represented as
$$
\zz{\alpha}{\beta}{\gamma}{\delta} = {1 \over 2} \Lambda_{\mu\nu} (\sigma^\mu)^\alpha_\beta (\sigma^\nu)^\gamma_\delta\,.
$$
Therefore, contraction of the first term inside brackets in (\ref{temperature.4}) gives
$$
2\cdot {1 \over 2} \Lambda_{\mu\nu} \sigma^\mu \cdot {K^\rho \over K^2}\cdot \mathrm{Tr}(\sigma^\nu \bar\sigma^\rho)
= 2 \Lambda_{\mu\nu} \sigma^\mu K^\nu\,,
$$
where for clarity we suppressed the remaining indices $\alpha$ and $\beta$, and we also used $K^2 = 1$.
To calculate the contraction of the second terms, we use identity
$$
{1 \over 2} (\sigma^\mu \bar\sigma^\rho \sigma^\nu + \sigma^\nu \bar\sigma^\rho \sigma^\mu) =
g^{\mu\rho}\sigma^\nu + g^{\nu\rho}\sigma^\mu - g^{\mu\nu}\sigma^\rho\,,
$$
which gives
$$
{1 \over 2} \Lambda_{\mu\nu} \cdot \sigma^\mu \bar\sigma^\rho \sigma^\nu \cdot K_\rho
 = {1 \over 2}(2\Lambda_{\mu\nu} - \mathrm{Tr}\Lambda \, g_{\mu\nu}) \sigma^\mu K^\nu\,,
$$
with $\mathrm{Tr}\Lambda = \Lambda_{\mu\nu} g^{\mu\nu} = \lambda_3-\lambda_4$.
Together, they make
$$
\delta {Y^\alpha}_\beta = - C (6\Lambda_{\mu\nu} - \mathrm{Tr}\Lambda \, g_{\mu\nu}) K^\nu {(\sigma^\mu)^\alpha}_\beta\,.
$$
Since $\delta {Y^\alpha}_\beta = \delta M^\mu (\sigma_\mu)^\alpha{}_\beta$, we finally get
\be
\delta M_\mu = - {T^2 \over 12}\left(6 \Lambda_{\mu\nu} - g_{\mu\nu}\mathrm{Tr}\Lambda\right) K^\nu\,.\label{temperature.6}
\ee

\end{document}